\documentclass[english,a4paper,oneside,11pt]{article}
\usepackage{babel}
\usepackage[dvips]{graphicx}

\newcommand{\titulo}[1]{\mbox{} \\ \noindent \textit{\textbf{\large #1}}\\}
\newcommand{\autor}[1]{\noindent \textbf{#1}}
\newcommand{\afil}[1]{{\small \noindent \textit{#1}}}

\renewcommand{\abstract}[1]{{\small \noindent \textbf{Abstract:} #1\\}}

\newcommand{\PACS}[1]{{\small \noindent \textbf{PACS numbers:} #1\\}}
\newcommand{\biblio}[1]{{\small \begin{thebibliography}{99} #1 \end{thebibliography}}}

\usepackage{bbm}
\usepackage{amsthm,amscd,amsmath,amssymb,amsfonts,here,epsf}
\begin{document}
 \thispagestyle{empty} 

\titulo{ River scaling analysis and the BHP universality hypothesis}




\autor{Rui Gon\c calves (email: rjasg@fe.up.pt)}

\afil{Faculdade de Engenharia do Porto}

\autor{Nico Stollenwerk (email: nks22@cus.cam.ac.uk)}

\afil{Instituto Gulbenkian de Ci\^encia}

\autor{Alberto Pinto (email: aapinto@fc.up.pt)}

\afil{Centro de Matem\'atica da Universidade do Porto}

\vspace{1.5cm}






\abstract{The XY-model shows in two dimensions in the strong
coupling regime a universal distribution, named BHP, which in turn also describes other models of criticality and self-organized criticality and even describes natural data as river level and flow.
We start by analysing the two dimensional XY-model and calculate the BHP probability density function. The results obtained for several dissimilar phenomenons which includes the deseasonalised Danube height data raised the universality hypothesis for rivers. This hypothesis is tested for the Iberian river Douro. Deviations from the BHP are found especially for medium and small runoffs. For regimes closer to the natural flow the fluctuations tend to follow the universal curve again.}



\section{Introduction}

In their pioneer paper the authors Bak, Tang and Wiesenfeld, \cite{BTW88}, proposed the hypothesis that under very general conditions nonequilibrium systems consisting of many interacting constituents may exhibit universal behaviour. This behaviour among other features is characterised by the formation of correlations and should occur naturally without any parameter tuning from the outside. For these systems the dynamical response is complex but the statistical properties are governed by simple power laws. Since only dimensionless numbers can be raised to arbitrary powers in a meaningful way, this self-organisation of the system implies independence of any particular scale of length or time. Hence the behaviour of the systems on these self-organized states is close to those of equilibrium systems at the critical point. This phenomenon is known as self-organized criticality (SOC).\\
In equilibrium thermodynamics criticality is linked with continuous phase transitions, \cite{Binneyetal}. The phase transition is continuous when the rate of change of some order parameter, and not the order parameter itself, is discontinuous at the critical point. In these systems when temperature is equal to the transition temperature the spin correlation function instead of an exponential decay reveals a power law behaviour and the corresponding exponent is called the critical exponent, in this case for the correlation function. As a consequence any local perturbation can be propagated through out the entire system and hence any member of the system affects all the others. At the critical point the contribution of the interaction between widely separated points for the large scale fluctuations of the order parameter are not exponentially rare. In this frame the central limit theorem is not applicable and Gaussianity is not present at all.\\
The existence of power laws statistics and critical exponents raises the question of classification of these systems. The phenomenon, whereby dissimilar systems exhibit the same critical exponents, is called universality. Two systems are assigned to the same universality class if they share the same dimensionality $d$ of the underlying lattice (number of spatial dimensions) and the same dimensionality of the order parameter, $D$. All the systems in the same universality class have the same critical exponents.\\
In this work we follow Bramwell, Holdsworth, Pinton, \cite{BHP1998} in calculating the probability density function (PDF) for the fluctuations of the magnetic order parameter of a two-dimensional model (2dXY) for spins in the strong coupling (low temperature) regime using the spin wave approximation where it shows a universal distribution, i.e., independent of system size and critical exponent $\nu$. This universal PDF describes other models of criticality and SOC \cite{bramwell2000} hence we show the data collapse for the deseasonalized Danube river height data at Nagymaros (80 years) following the steps of \cite{DahlstedtJensen2005} and compare it with the daily mean runoff data of river Douro at R\'egua (26 years). According to those authors no important differences occur when height or runoff measures are used.\\
This article is organised as follows the second section describes the two-dimensio\-nal Ising model (2dIsing) and its frame set as the preliminary for the 2dXY calculations presenting, order parameter and PDF's at different regimes of temperature; subcritical critical and supercritical. In the third section we present the analytics for the spin-wave (SW) approximation of the 2dXY model and the magnetic order parameter PDF for the critical (low temperature) regime named BHP, after Bramwell, Holtsworth and Pinton \cite{bramwell2000}. In the fourth section we show the data collapse of the deseasonalized Danube river height data and we compare it with the Douro river in two cases, the entire year and the winter regime. Apparently the deviations of the Douro river from the BHP are due to some river flow regulations in order to have fluctuations closer to the Gaussian PDF. For larger values of the streamflow the regulation is no longer possible due to storage limitations and the river dams are set open. For larger values the river fluctuations are much closer to the BHP form. The main differences between the Danube and Douro are related to the amount of water on the river basin and its distribution in time. The Douro is a southern European river where the summer is usually very dry contrasting with the Danube basin where there is lots of water the year around and regulation seems not to have any major effect on the natural flow.

\section{The Ising model}

Ernst Ising suggested a very simple thermodynamic model 
to understand spontaneous magnetization
(Ising, 1925, \cite{Ising}). Statistical mechanics as 
used by Ising considers equilibrium distributions
\begin{equation}
        p^*(\sigma _1, ... , \sigma _N)=\frac{1}{Z}
                   \cdots e^{\cal{ H} (\sigma _1, ... , \sigma _N)}
        \label{pequilibrium}
\end{equation}
\noindent
and quantities derived from them, like the 
magnetization per spin $\langle M \rangle := \langle \frac{1}{N} 
\sum _{i=1}^{N} \sigma _i \rangle $ 
and magnetic
susceptibility $\chi := \langle M^2 \rangle - \langle M \rangle ^2$.
Here the $N$ spins $\{ \sigma _i \}_{i=1}^{N}$, $\sigma _i \in \{-1,+1\} $, 
are distributed 
in $p^*(\sigma _1, ... , \sigma _N)$ according to 
Boltzmann weights $e^{\cal H} $ (originally
introduced to represent well known distributions 
in ideal gas theory, the Maxwell distribution
of velocities) with a function 
\begin{equation}
        {\cal H} (\sigma _1, ... , \sigma _N) =
                \sum_{i=1 }^{N} \sum_{j=1 }^{N} V_{ij} \sigma _i \sigma _j + 
                \sum_{i=1 }^{N} h_i \sigma _i
                +\sum_{i=1 }^{N} c_i
        \label{hamiltonian}
\end{equation}
\noindent
and a normalization, the partition function $Z$. 
In order to obtain reasonably sized numbers we
use $C=\sum_{i=1 }^{N} c_i =-N \mbox{ln} 2$ for the $2^N$ possible 
configurations of spins up and down, hence
\begin{equation}
        Z=
                \sum_{\sigma _1=\pm 1 } ... \sum_{\sigma _N=\pm 1 } 
                e^{
                \sum_{i=1 }^{N} \sum_{j=1 }^{N} V_{ij} \sigma _i \sigma _j + 
                \sum_{i=1 }^{N} h_i \sigma _i
                }
                \cdot \frac{1}{2^N}
                \quad.
        \label{partitionfunction}
\end{equation}
\noindent
It is $V_{ij}:=V \cdot J_{ij}$ with coupling strength $V$ 
and adjacency matrix $J_{ij} \in \{0,1 \}$
and $h:=h_i$ the external magnetic field.
Then the distribution of the magnetization per spin is given by
\begin{equation}
    p(M)=
        \sum_{\sigma _1=\pm 1 } ... \sum_{\sigma _N=\pm 1 } 
                \delta \left( M -  \frac{1}{N} \sum _{i=1}^{N} \sigma _i  \right)
                p^*(\sigma _1, ... , \sigma _N)
        \label{magdistribution}
\end{equation}
\noindent
from simply applying Bayes' rule 
$p(M)= \sum_{\sigma _1=\pm 1 } ... 
    \sum_{\sigma _N=\pm 1 } p(M|\sigma _1, ... , \sigma _N)
\cdot p(\sigma _1, ... , \sigma _N)$ for joint and conditional probabilities.

The mean magnetization, $\langle M \rangle=\sum_{i=0}^{N} M \cdot p(M)$, with $M=-1+i\cdot \Delta M$, $\Delta M =2/N$ 
is given by
\begin{equation}
        \langle M \rangle 
               =\sum_{\sigma _1=\pm 1 } ... \sum_{\sigma _N=\pm 1 } 
                \left(
                \sum_{i=0}^{N} M \cdot
                        \delta \left( M -  \frac{1}{N} 
                          \sum _{i=1}^{N} \sigma _i  \right)
                \right)
                        p^*(\sigma _1, ... , \sigma _N)
        \quad.
        \label{mag}
\end{equation}
\noindent
Using the $\delta $-function the mean magnetization is then
\begin{equation}
        \langle M \rangle 
        = 
                \sum_{\sigma _1=\pm 1 } ... \sum_{\sigma _N=\pm 1 } 
                \left(
                 \frac{1}{N} \sum _{i=1}^{N} \sigma _i  
                \right)
                        \frac{1}{Z} e^{\cal H}
        \quad.
        \label{mag2}
\end{equation}
\noindent
The mean magnetization $\langle M \rangle $ 
is an easily accessible quantity in physical systems,
whereas the distribution of the magnetization $p(M)$ 
is easily accessible in small systems.
Also the mean $\langle M \rangle $ is dominated 
by the maximum $M_{max}$ of the distribution
$p(M)$ for large systems.
The Ising model was proposed by Lenz\cite{Lenz} and solved by Ising for one dimension. The two dimensional case shows already a phase transition. In 1944 Onsager\cite{Onsager}, solved the Ising system for $d=2$ in the absence of an external magnetic field. Here we present a numerical simulation for the magnetization distribution for the 2d Ising model for a $N=4\times4$ lattice with different coupling strength, $V$. The external magnetic field is absent, $h=0$.

\begin{figure}[!htb]
\begin{minipage}[t]{0.30\linewidth}
\includegraphics[width=\linewidth]{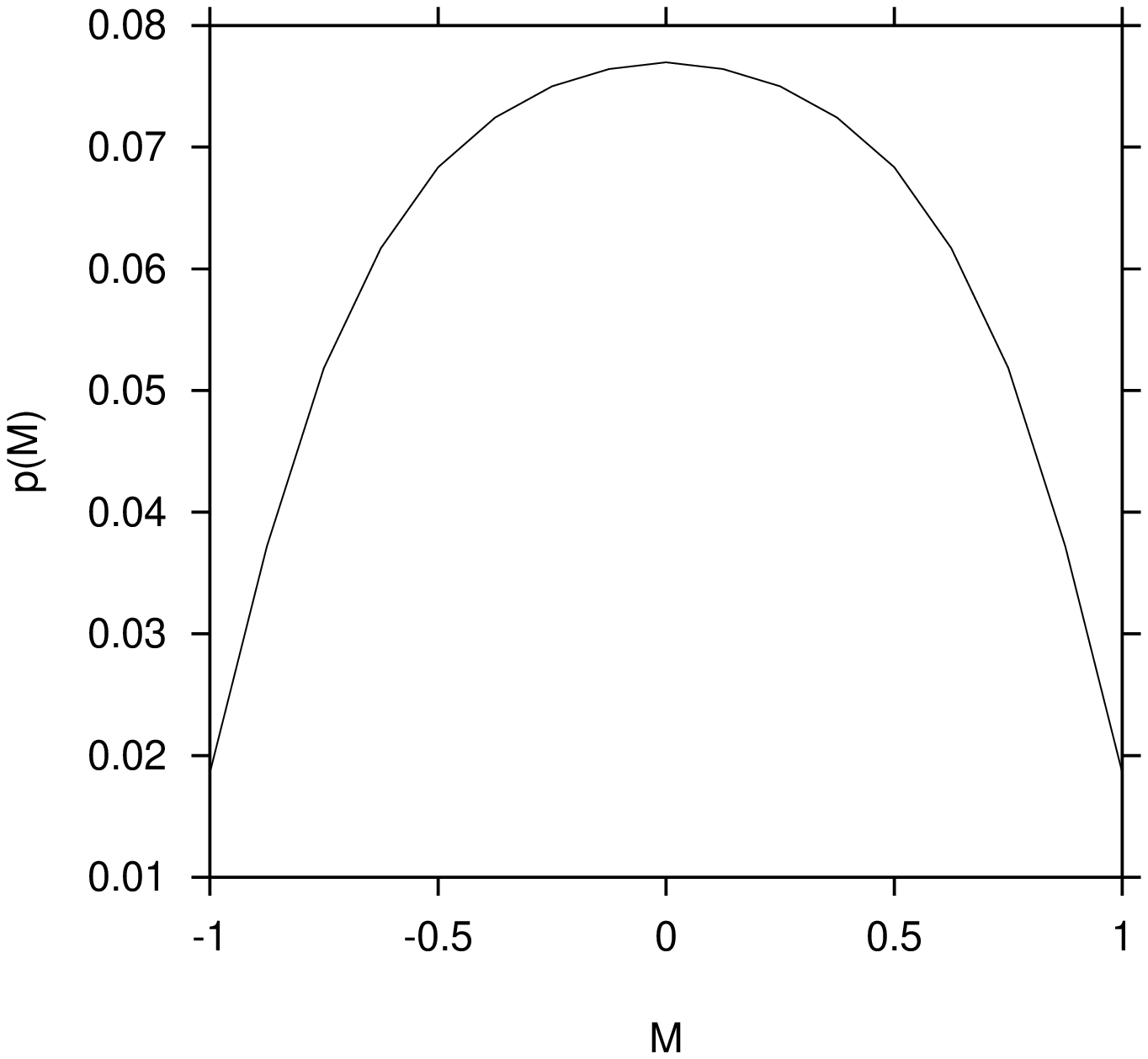}
\caption{\footnotesize{Coupling strenght value is $V=0.13$. The system is subcritical, 
only one maximum at $M=0$ appears.}}
\label{fig:vertroh0130}
\end{minipage} \hfill
\begin{minipage}[t]{0.30\linewidth}
\includegraphics[width=\linewidth]{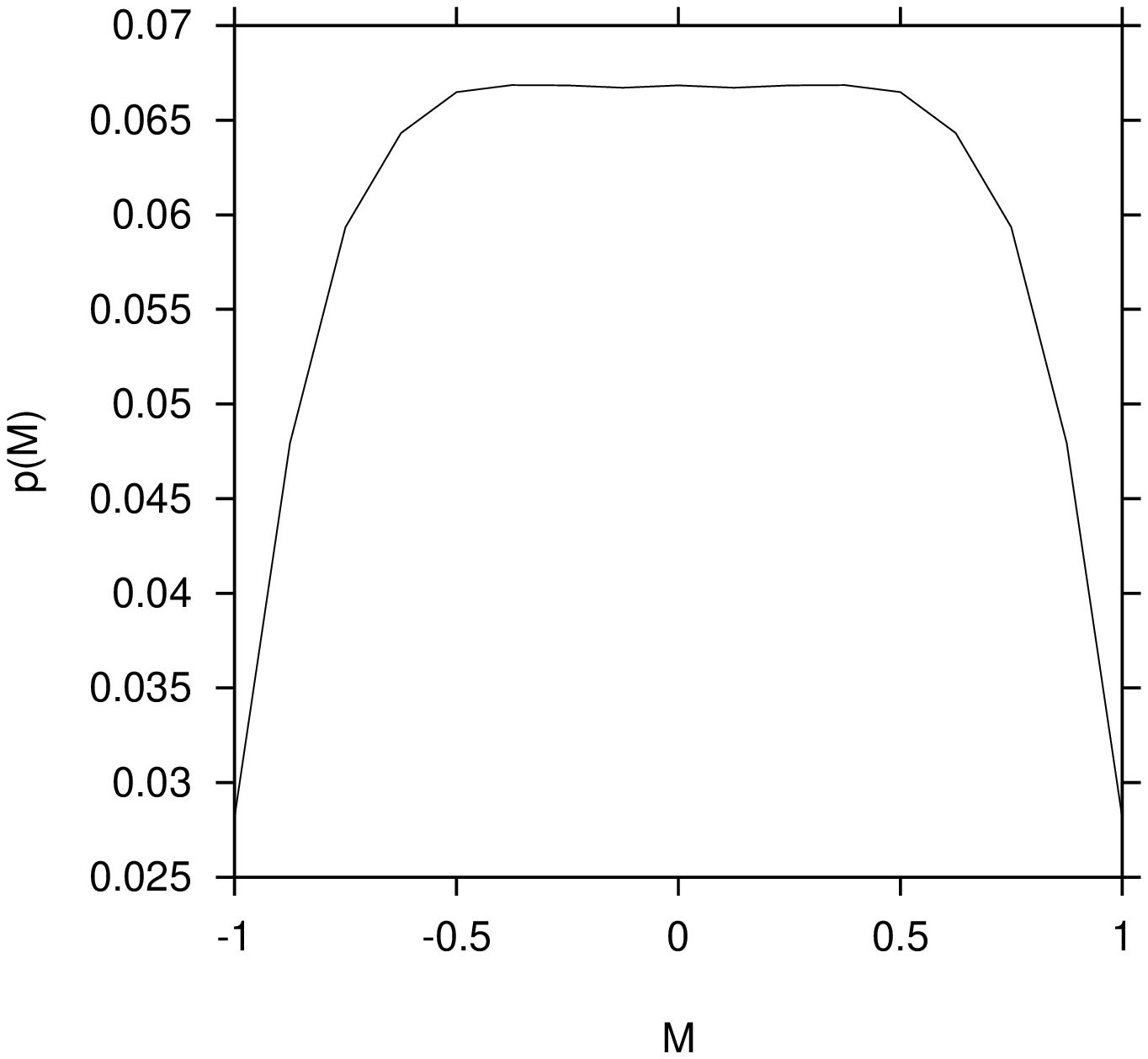}
\caption{\footnotesize{ For the value of $V=0.14\approx V_c$ we find one maximum at $M=0$, but being broadened up
just before splitting into the two symmetric maxima $M_{max} \neq 0$.}}
\label{fig:vertroh0140}
\end{minipage} \hfill
\begin{minipage}[t]{0.30\linewidth}
\includegraphics[width=\linewidth]{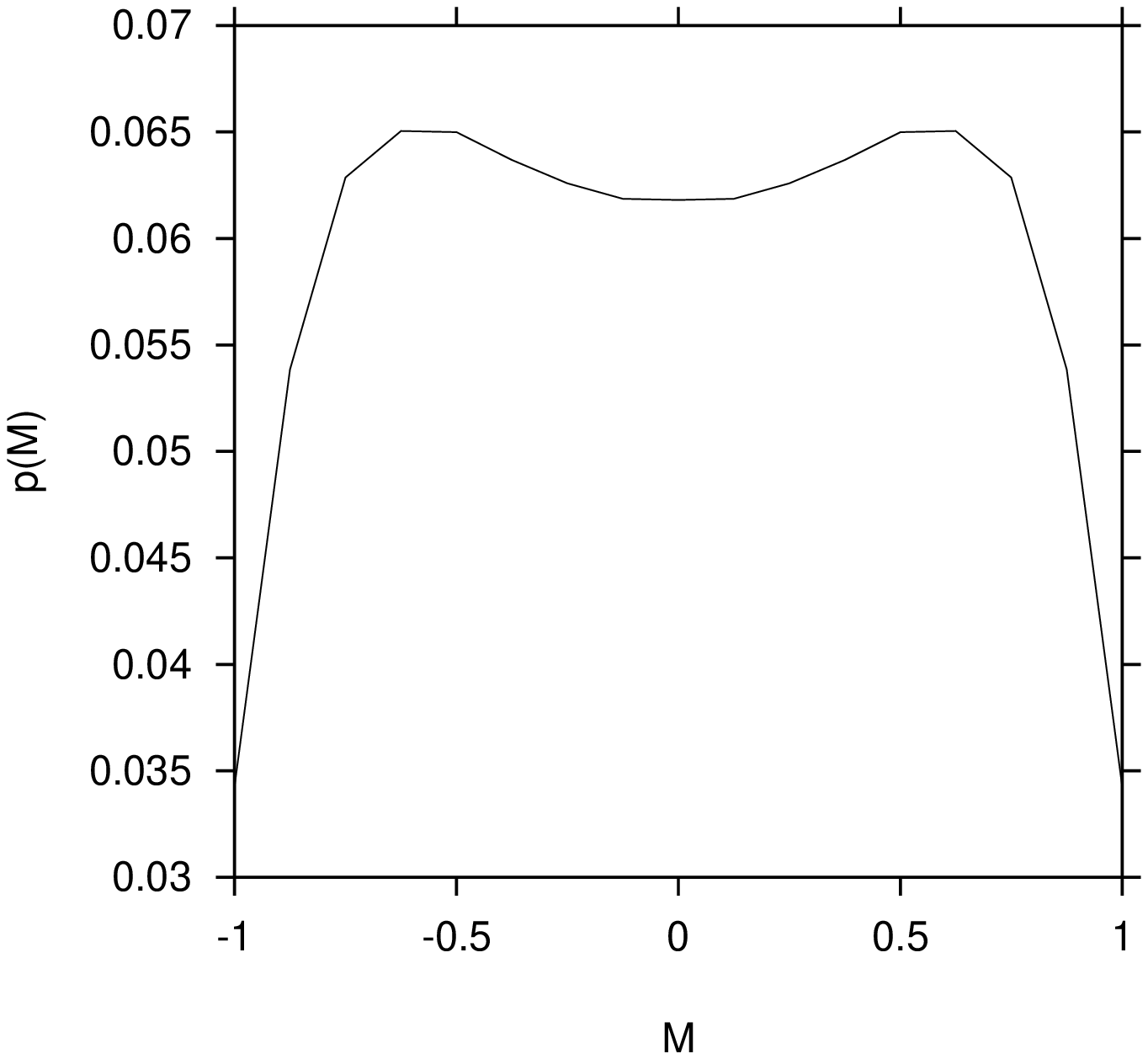}
\caption{\footnotesize{ Coupling strenght value is $V=0.145$ two maxima $M_{max} \neq 0$ are found symmetrically 
around $M=0$, where there is a local minimum in $p(M)$.}}
\label{fig:vertroh0145}
\end{minipage}       
\end{figure}
\noindent
\normalsize
For low temperature (strong coupling), fig. \ref{fig:vertroh0130} the spins are strongly correlated and tend to align with their neighbours so the distibution mode is for zero magnetization. Near critical temperature, fig. \ref{fig:vertroh0140} small variations of spins can be propagated through the lattice and will affect spins at large distances. That is the point of phase transition. Fluctuations can easily happen for values 
between $M\approx -0.5$ and $M\approx +0.5$, for which all
magnetizations in this range are about equally likely. For temperatures just above criticality, fig. \ref{fig:vertroh0145}, two maxima start appearing symmetrically around the origin.

\subsection{The XY-model in two dimensions}

In the XY model spins are able to rotate freely in the XY-plane, i.e. the
spin variables are $\underline S_i$ with
\begin{equation}
  \underline S = \left(
                   \begin{array}{c}
                     cos(\theta _i)   \\
                     sin(\theta _i)   \\
                   \end{array}
                 \right)
        \quad
        \label{xyspins}
\end{equation}
\noindent
where the angle $\theta _i $ changes between $-\pi$ and $+\pi$.\\
The stationary distribution is again given by
\begin{equation}
        p^*(\underline S _1, ... ,\underline S  _N)=\frac{1}{Z}
           \cdot e^{{\cal H} (\underline S _1, ... ,\underline S  _N)}
        \label{pxyequilibrium}
\end{equation}
\noindent
and 
\begin{equation}
        {\cal H} (\underline S _1, ... ,\underline S  _N) =
                \sum_{i=1 }^{N} \sum_{j=1 }^{N} V_{ij} 
                      \underline S_i \underline S _j + 
                \sum_{i=1 }^{N} \underline H \, \underline S _i
                +C
        \label{xyhamiltonian}
\end{equation}
\noindent
where later the constant $C$ will be fixed to obtain reasonably
small values for the partition function $Z$
(irrelevant for all physically relevant quantities, e.g.
the stationary distribution $p^* $).

Since the spins are unit vectors rotating we can replace them just by
angles between two interacting spins
\begin{equation}
       \underline S_i \underline S _j 
            = |S_i| \cdot |S_j| \cdot  cos (\theta _i -\theta _j)
            = cos (\theta _i -\theta _j)
        \label{xythetainteraction}
\end{equation}
\noindent
or the angle between spin and outer magnetic field (see \cite{Binneyetal, ZinnJustin})
\begin{equation}
       \underline H \; \underline S _i 
            = |H| \cdot |S_i| \cdot  cos (\theta _i )
            = h \cdot cos (\theta _i )\quad.
        \label{xythetamagneticfield}
\end{equation}
\noindent
Hence
\begin{equation}
        p^*(\theta _1, ... ,\theta   _N)=\frac{1}{Z}
           \cdot e^{{\cal H} (\theta _1, ... ,\theta   _N)}
        \label{pxyequilibriumtheta}
\end{equation}
\noindent
and 
\begin{equation}
        {\cal H} (\theta  _1, ... \theta   _N) =
                V \sum_{i=1 }^{N} \sum_{j=1 }^{N} J_{ij} 
                      cos( \theta _i - \theta _j) + 
                \sum_{i=1 }^{N} h \cdot cos(\theta _i)
                +C
        \label{xyhamiltoniantheta}
\end{equation}
\noindent
with partition function as normalization constant
\begin{equation}
        Z=\int_{-\pi }^{\pi} \, d\theta _1 ... \int_{-\pi }^{\pi} \, d\theta _N
              \; \; e^{{\cal H} (\theta _1, ... ,\theta   _N)}\quad.
        \label{xypartitionfct}
\end{equation}
\noindent
The absolute value of the magnetization as order parameter for the
phase transition \cite{KosterlitzThouless} is given by

\begin{equation}
        m:= \frac{1}{N}  \parallel \sum_{i=1}^{N} \underline S_i 
                         \parallel        
            = \frac{1}{N} \sqrt{\left(
                                   \sum_{i=1}^{N} cos(\theta _i)
                                 \right)^2
                               +
                                 \left(
                                   \sum_{i=1}^{N} sin(\theta _i)
                                 \right)^2
                             }
            =: M(\theta _1, ... ,\theta   _N)
        \label{xymagnetization}
\end{equation}
\noindent
and the distribution of this quantity
\begin{equation}
        p(m)=\int_{-\pi }^{\pi} \, d\theta _1 ... \int_{-\pi }^{\pi} \, d\theta _N
              \; \;  \delta (m-M(\theta _1, ... ,\theta   _N)  )
              \; \;\frac{1}{Z} e^{{\cal H} (\theta _1, ... ,\theta   _N)}\quad.
        \label{xydistribmagnetization}
\end{equation}
\noindent
This defines completely the XY-model in any dimension (where the dimension
is fixed by specifying the adjacency matrix $J$ in 
${\cal H} (\theta _1, ... ,\theta   _N) $). This will be referred to the 
non-quadratic model, since later for analytical treatability of the
strong coupling regime the $cos$-interaction will be replaced by the
quadratic Taylor expansion (valid for small angles between the spins only
or spins and external magnetic field).
The BHP-distribution will turn up by evaluating 
the Eq. (\ref{xydistribmagnetization}) in the case of quadratic 
approximation of the above model.
Using the second order Taylor series expansion for the cosine,
\begin{equation}
        cos(x)=1-\frac{1}{2} x^2 + {\cal O} (x^4)
        \label{cosTaylor}
\end{equation}
\noindent
hence we can replace in ${\cal H} (\theta _1, ... ,\theta   _N) $
in the strong coupling case ($V>>1$)
\begin{equation}
        cos(\theta _i - \theta _j)
            \approx 1-\frac{1}{2} (\theta _i - \theta _j)^2 
        \label{cosTaylortheta}
\end{equation}
\noindent
i.e. approximating all relevant
quantities in quadratic form in the exponential function and then
use Gaussian integration. This approximation is valid for strong
coupling $V>>1$, where in the physical model spin waves are the
only relevant dynamical phenomena.\\
The partition function as normalization constant
\begin{equation}
        Z=\int_{-\pi }^{\pi} \, d\theta _1 ... \int_{-\pi }^{\pi} \, d\theta _N
              \; \; e^{{\cal H} (\theta _1, ... ,\theta   _N)}
        \label{xypartitionfct2}
\end{equation}
\noindent
becomes in quadratic approximation
\begin{equation}
        Z=\int_{-\infty }^{\infty  } \, d\theta _1 ... 
              \int_{-\infty  }^{\infty } \, d\theta _{N-1} ...
              \int_{-\pi }^{\pi} \, d\theta _N
              \; \; e^{J_{tot}\cdot V}
              \; \; e^{-V \cdot \underline{\theta} ^{tr} A \underline{\theta} }
        \label{xypartitionfctquadratic}
\end{equation}
\noindent
It is
$A:= Q\cdot \mathbbm{1} -J$ with $J$ the adjacency matrix, 
$Q:=  \sum_{j=1 }^{N} J_{ij}$
the number of neighbours, and $J_{tot}:= \sum_{i=1 }^{N} \sum_{j=1 }^{N} J_{ij}$
the total number of connections in the adjacency matrix.
Changing variables $\underline y = T \underline{\theta}$ with the 
transformation matrix
\begin{equation}
  T := ( u_{nk}) = \left(
                       \frac{1}{\sqrt{N}}
                       e^{-2 \pi i \frac{1}{L} \underline k \cdot \underline n}
                  \right)
        \label{fouriertransform}
\end{equation}
\noindent
diagonalizes matrix $A$ such that $T^{\dagger} AT = \Lambda $ with
$ \Lambda $ the diagonal matrix of eigenvalues of $A$ and $\underline u _k$
the eigenvectors of $A$ due to the structure of the adjacency matrix $J$.
$T^{\dagger}  $ is the Hermitean matrix of the complex matrix $T$,
i.e. the transposed and complex conjugate.
In 2 dimensions 
\begin{equation}
   \underline k \cdot \underline n
                  =
                  \left(
                   \begin{array}{c}
                     k_x \\
                     k_y \\
                   \end{array}
                 \right)
                                  \cdot
                  \left(
                   \begin{array}{c}
                     n_x \\
                     n_y \\
                   \end{array}
                 \right)
        \label{fouriertransformvectors}
\end{equation}
\noindent
with $k_x$ etc. having values 1,2,..,L, and with $N=L\times L$, the variable
\begin{equation}
   k:=k_x + (k_y-1) \cdot L
        \label{fouriertransformbetraege}
\end{equation}
\noindent
has values 1,2,..., N. Likewise for $\underline n$.
Also in 2d the eigenvalues are
\begin{equation}
   \lambda_k =
        4-2 \cdot \cos \left( 2 \pi \frac{1}{L}  k_x \right)
        -2 \cdot \cos \left( 2 \pi \frac{1}{L}  k_y \right)\quad.
        \label{eigenwertevonA}
\end{equation}
\noindent
In the general case the partition function is then given 
in changed coordinates, the eigencoordinates of $A$,
\begin{equation}
        Z=  e^{J_{tot}\cdot V}
            \int_{-\infty }^{\infty  } \, dy _1 ... 
              \int_{-\infty  }^{\infty } \, dy _{N-1 } ...
              \int_{-c }^{c} \, dy _N
              \; \; e^{-V \cdot \underline{y} ^{tr} \Lambda \underline{y} }
              \cdot |det(T)|
        \label{xypartitionfcteigencoordinates}
\end{equation}
\noindent
with $|det(T)|=1$ and transformed integration boundaries for the
zero eigenvalue coordinates $c$. It turns out that the $N^{th}$ coordinate
is the zero eigenvalue coordinate, in which case
$c= (u^{\dagger }_{NN})^{-1} \pi = \sqrt{N} \cdot \pi$.

\begin{figure}[h!]
\begin{center}
\includegraphics*[width=8cm]{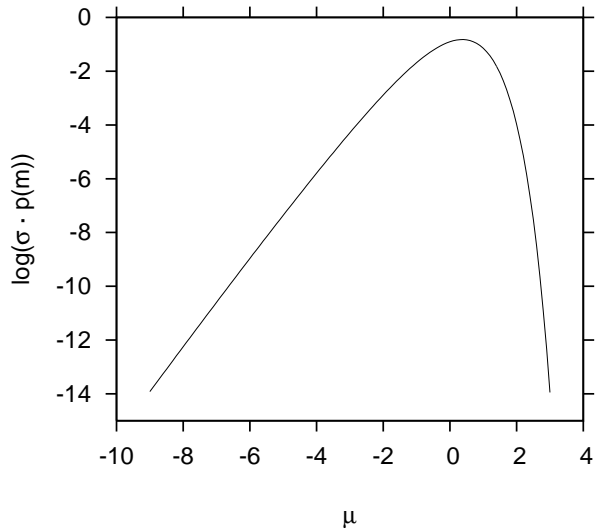}
\caption{\footnotesize{Logarithm of the universal BHP PDF for the 2dXY model with $L=10$.}} \label{fig:vertpm12}
\end{center}
\end{figure}

Performing the Gaussian integration\footnote{Let $B$ be a complex symmetric matrix with a non-negative real part and non zero eigenvalues, then, 
$             \int_{-\infty}^{\infty}\cdots\int_{-\infty}^{\infty}d\underline{\theta}\,e^{-\underline{\theta}^T B\underline{\theta}}=\sqrt{\frac{\pi^N}{\det{B}}}$.}
of 
Eq. \ref{xypartitionfcteigencoordinates} gives
\begin{equation}
        Z=  e^{J_{tot}\cdot V}
            \prod _{k=1}^{N-1 }
            \sqrt{\frac{\pi}{V\cdot \lambda _k}}
            \cdot (2\sqrt{N}\pi)
        \label{xypartitionfctgaussintegration}
\end{equation}
\noindent
where the last factor comes from the eigenvalue zero. Using the Fourier representation of the delta function in \ref{xydistribmagnetization}, we obtain
 
\begin{equation}
     p(m)=\int_{-\infty}^{\infty}\cdots\int_{-\infty}^{\infty}\int_{-\pi }^{\pi} \, d\theta _1\cdots  d\theta _N
          \int_{-\infty}^{\infty}\frac{1}{2\pi}e^{i(m-M(\underline{\theta})x}dx\frac{1}{Z} e^{{\cal H} (\theta _1, \cdots ,\theta   _N)}\quad.
    \label{xydistribmagnetization3}
\end{equation}
\noindent
For simplicity purposes it's more convenient to write the magnetization in terms of the spin variables instead of vector spins, 

\begin{equation}
  M=\frac{1}{N}\sum_{r=1}^N(\theta_r-\overline{\theta})^2
    \label{magnetizationabsolute}
\end{equation}
\noindent
this is just the average, $\langle\theta_{r}\rangle$ for unconfined spins. For the case of periodic spins, following \cite{Archambault1997}, the instantaneous magnetization direction, $\overline{\theta}$, is defined by,

\begin{equation}
 \overline{\theta} ={\it arctan} \left( {\frac {\sum _{r=1}^{N}\sin{{\theta_r }}}{\sum _{r=1}^{N}\cos{{\theta_r }}}}
 \right)
\end{equation}
\noindent
which is the same as the average for large $N$. Then we may write $M({\underline{\theta}})= 1-\frac{1}{2N}\underline{\theta}^{tr}\underline{\theta}$. The magnetization PDF can be written using the quadratic approximation,

\begin{equation}
              p(m)=\frac{e^{J_{tot}}}{Z}\int_{-\infty}^\infty \frac{dx}{2\pi}\int_{-\infty}^\infty \cdots\int_{-\infty}^\infty\int_{-\pi}^{\pi} d\underline{\theta}e^{ix\left(m-1+\frac{1}{2N}\underline{\theta}^{tr}\underline{\theta}\right)-V\underline{\theta}^{tr}A\underline{\theta}}
        \label{xydistribmagnetization4}
\end{equation}
\noindent
where $d\underline{\theta}=\prod_{k=1}^N d\theta_k$. Simplifying for Gaussian integration,

\begin{equation}
              p(m)=\frac{e^{J_{tot}}}{Z}\frac{1}{2\pi}\int_{-\infty}^\infty \frac{dx}{2\pi}\int_{-\infty}^\infty \cdots\int_{-\infty}^\infty\int_{-\pi}^{\pi} d\underline{\theta}e^{\underline{\theta}^{tr}\left(\frac{ix}{2N}\mathbbm{1}-VA\right)\underline{\theta}}\quad.
        \label{xydistribmagnetization5}
\end{equation}
\noindent
Using the linear transformation (\ref{fouriertransform}) and writing $B=VA-\frac{ix}{2N}\mathbbm{1}$ there is a diagonal  matrix $\tilde{\Lambda}$ such that $B=T\tilde{\Lambda} T^{\dagger}$. Using the multivariate Gaussian integral, we obtain

\begin{equation}
              \int_{-\infty}^{\infty}\cdots\int_{-\infty}^{\infty}d\underline{\theta}e^{-\underline{\theta}^T B\underline{\theta}}=\sqrt{\frac{\pi^N}{\det{B}}}\quad.
        \label{multivariategaussianintegral}
\end{equation}
\noindent
The matrix $B$ is equal to $q(A)$ where $q()$ is a polynomial of $A$. Then we may write,

\begin{equation} 
 \det(B)=\prod_{k=1}^N \ell_k=\prod_{k=1}^N q\left(\lambda_k\right)
        \label{determinantofB}
\end{equation}
\noindent
where $\left\{\lambda_k\right\}_{k=1}^N$ are the eigenvalues of matrix $A$ and $\left\{\ell_k\right\}_{k=1}^N$ are the eigenvalues of $B$. The magnetization PDF is,

\begin{equation}
              p(m)=\frac{(2c)e^{J_{tot}}}{Z}\int_{-\infty}^{\infty}\frac{dx}{2\pi}e^{ix(m-1)}\sqrt{\frac{\pi^{N-1}}{\prod_{k=1}^{N-1} \ell_k}}
        \label{xydistribmagnetization6}
\end{equation}
\noindent
where $c=\sqrt{N}\pi$. Simplifying we have,

\begin{equation}
              p(m)=\int_{-\infty}^\infty\frac{dx}{2\pi}e^{ix(m-1)}\left(\prod_{k=1}^{N-1}\frac{ \ell_k}{V\lambda_k}\right)^{-1/2}\quad.
        \label{xydistribmagnetization7}
\end{equation}
\noindent
The mean and standard deviation magnetization can be calculated using the distribution of the magnetization absolute value, \ref{xydistribmagnetization6}. Under the same assumption for the $A$ matrix we have

\begin{eqnarray}
\left\lbrace \begin{array}{c}
\langle m\rangle=1-\frac{1}{2N}\sum_{k=1}^{N-1}\frac{1}{2V\lambda_k}\\
\\
\sigma_m^2=\frac{1}{2N}\sum_{k=1}^{N-1}\frac{1}{2}\left(\frac{1}{N\lambda_k}\right)^2\quad.
\end{array}
\right.
    \label{BHPmoments}
\end{eqnarray}
\noindent
At this point one has to do the appropriate normalisation in order to get the universal critical PDF. The distribution function for the magnetization has been discussed by several authors (see references in \cite{Archambault1997}). Binder, 1992\cite{Binder}, argued that from the Ising model with distribution $f(m)$ there is scaling of the form

\begin{equation}
              p(m)\sim L^{\beta/\nu}p(mL^{\beta/\nu},L/\zeta)
        \label{BHPscaling1}
\end{equation}
\noindent
where $\zeta$ is the correlation length. We know that very close to the critical point the correlation length is going to be much bigger than the system size. In this case $L$ is the only important scale so $f$ should depend only on the variable $mL^{\beta/\nu}$,

\begin{equation}
              p(m)\sim L^{\beta/\nu}p(mL^{\beta/\nu})\quad.
        \label{BHPscaling2}
\end{equation}
\noindent
This form is not surprising since that for this system the mean $\langle m \rangle$ scales with $L^{-\beta/\nu}$ and being critical the standard deviation, $\sigma$ is also scales in the same way as the mean value. These facts explain the universal form of $p(mL^{\beta/\nu})$. Hence we can rewrite equation (\ref{BHPscaling2}) as

\begin{equation}
              p(m)\sim \frac{1}{\sigma_m}f(\frac{m}{\sigma_m})\quad.
        \label{BHPscaling3}
\end{equation}
\noindent
The standard deviation is the correct normalization for the order parameter. Defining $\mu$ as

\begin{equation}
              \mu=\frac{m-\langle m\rangle}{\sigma_m}
        \label{normalizedm}
\end{equation}
\noindent
it is straightforward to see that the PDF $p(\mu)=\sigma_mp(m)$ and this PDF is normalized not depending on system size

\begin{equation}
          \sigma _m \cdot p(m) \sim  L^0 \cdot 
f\left( \frac{m-\langle m \rangle}{\sigma _m}  \cdot L^0 \right)\quad.
        \label{BHPscaling4}
\end{equation}
\noindent
The form of the normalized and consequently universal BHP is,

\begin{eqnarray}
              p(\mu)=\int_{-\infty}^{\infty}\frac{dx}{2\pi} 
\sqrt{\frac{1}{2N^2}\sum_{k=1}^{N-1}\frac{1}{\lambda_k^2}}      
&\!&e^{ix\mu\sqrt{\frac{1}{2N^2}\sum_{k=1}^{N-1}\frac{1}{\lambda_k^2}}
-\sum_{k=1}^{N-1}\left[\frac{ix}{2N}\frac{1}{\lambda_k}-\frac{i}{2}
\mbox{arctan}\left(\frac{x}{N\lambda_k}\right)\right]}\nonumber\\\!
&\!&.e^{-\sum_{k=1}^{N-1}\left[\frac{1}{4}\mbox{ln}{\left(1+\frac{x^2}{N^2\lambda_k^2}\right)}\right]}
\quad.
      \label{BHPfinal}
\end{eqnarray}
\noindent
In figure \ref{fig:vertpm12} we show a BHP PDF simulation for $L=10$, $N=L^2$, using eq. (\ref{BHPfinal}).

\noindent
At this point we should mention that the magnetic order parameter density of the 2dXY model approaches this universal PDF when $T\rightarrow0$. Surprisingly the quadratic approximation remains close to the true PDF for almost all the critical regime. It has been shown recently \cite{MackPalmaVergara2005, BanksBramwell} that the weak temperature dependence gets noticeable near the KT phase transition.


\section{BHP and the Danube and Douro data}


J\'anosi and Gallas \cite{ImreJason99} analysed statistics of the daily water level fluctuations of the Danube collect over the period 1901-97 at Nagymaros, Hungary. The authors found in the Danube data similar characteristics to those of company growth\cite{Stanleyetal}. This suggested that a universal description of the statistics should exist for both systems. Those authors noticed a data collapse for the conditional PDF of the one day logarithmic rate of changes for the mean adjusted river height.\\
Following \cite{ImreJason99}, let $h(t)$ be the 87$\times$365 data points\footnote{all the $29^{\mbox{th}}$ of February were discarded} the time series of height measurements and define a season mean and, following \cite{bramwellfennelleurophys2002}, a season standard deviation, $\overline{h}(t)$ and $\sigma_h$, respectively,

\begin{equation}
          \overline{h}(t)=\frac{1}{87}\sum_{j=0}^{86} h(t,j) 
        \label{seasonadjustment1}
\end{equation}
\noindent
where $h(t,j)=\left[\left\lbrace h(t+j*365)\right\rbrace_{j=0}^{86}\right] _{t=1}^{365}$,

\begin{equation}
          \sigma_h(t)=\sqrt{\frac{\sum_{j=0}^{86} h(t,j)^2}{87}-{\overline{h}(t)}^2}\quad.
        \label{seasonadjustment2}
\end{equation}
\noindent
The seasonality adjustment consists in normalizing each value of the original time series by subtracting the respective component of the season mean and dividing by the respective component of the season standard deviation, $\left(\frac{h-\overline{h}}{\sigma_h}\right) $. The water plays for the river the same role as the coupling strength in the 2dXY model.
\begin{figure}[!htb]
\begin{minipage}[t]{0.45\linewidth}
\includegraphics[width=\linewidth]{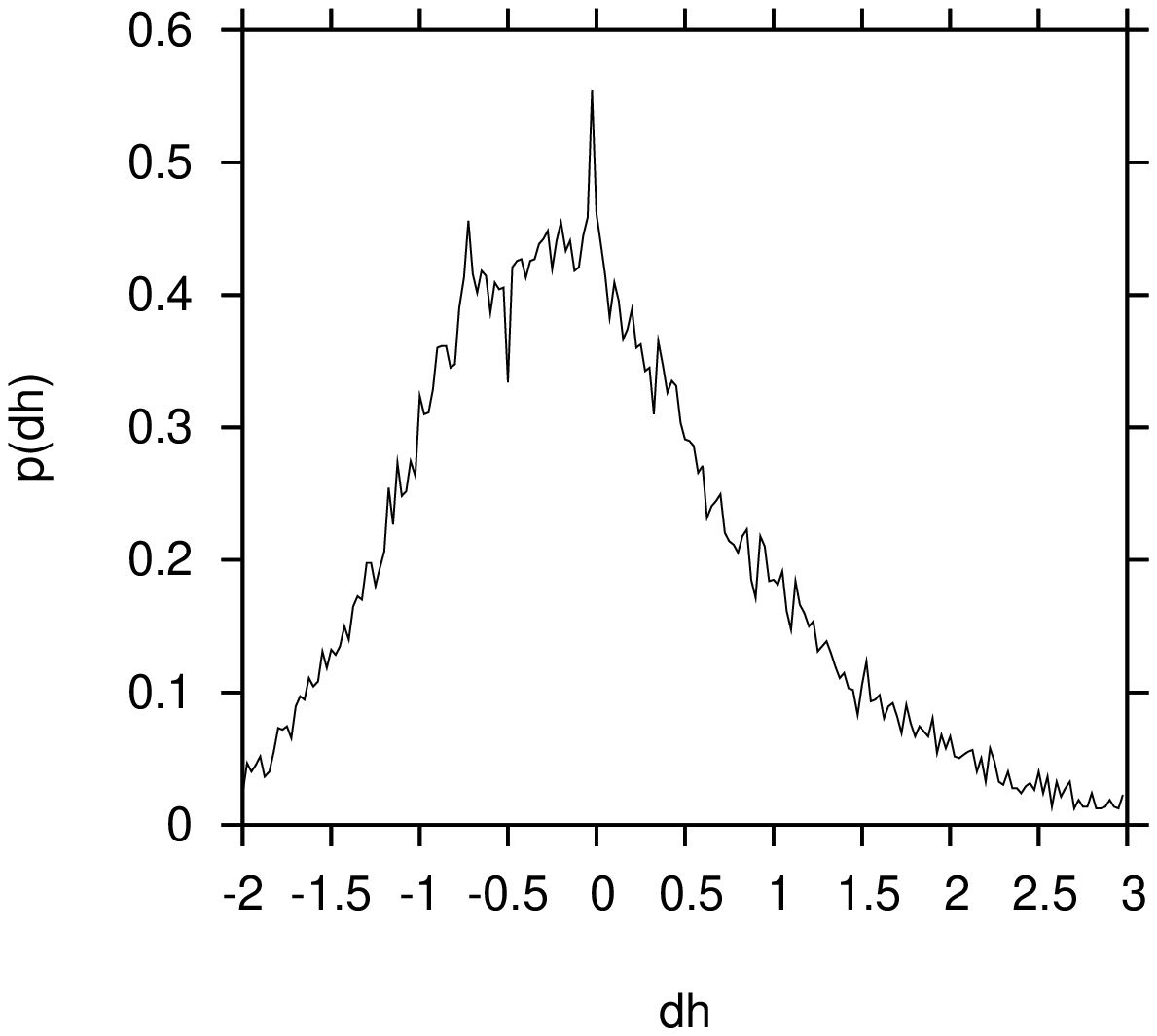}
\caption{\footnotesize{Histogram of the seasonally adjusted daily water height Danube river data.}}
\label{fig:donaubin3}
\end{minipage} \hfill
\begin{minipage}[t]{0.45\linewidth}
\includegraphics[width=\linewidth]{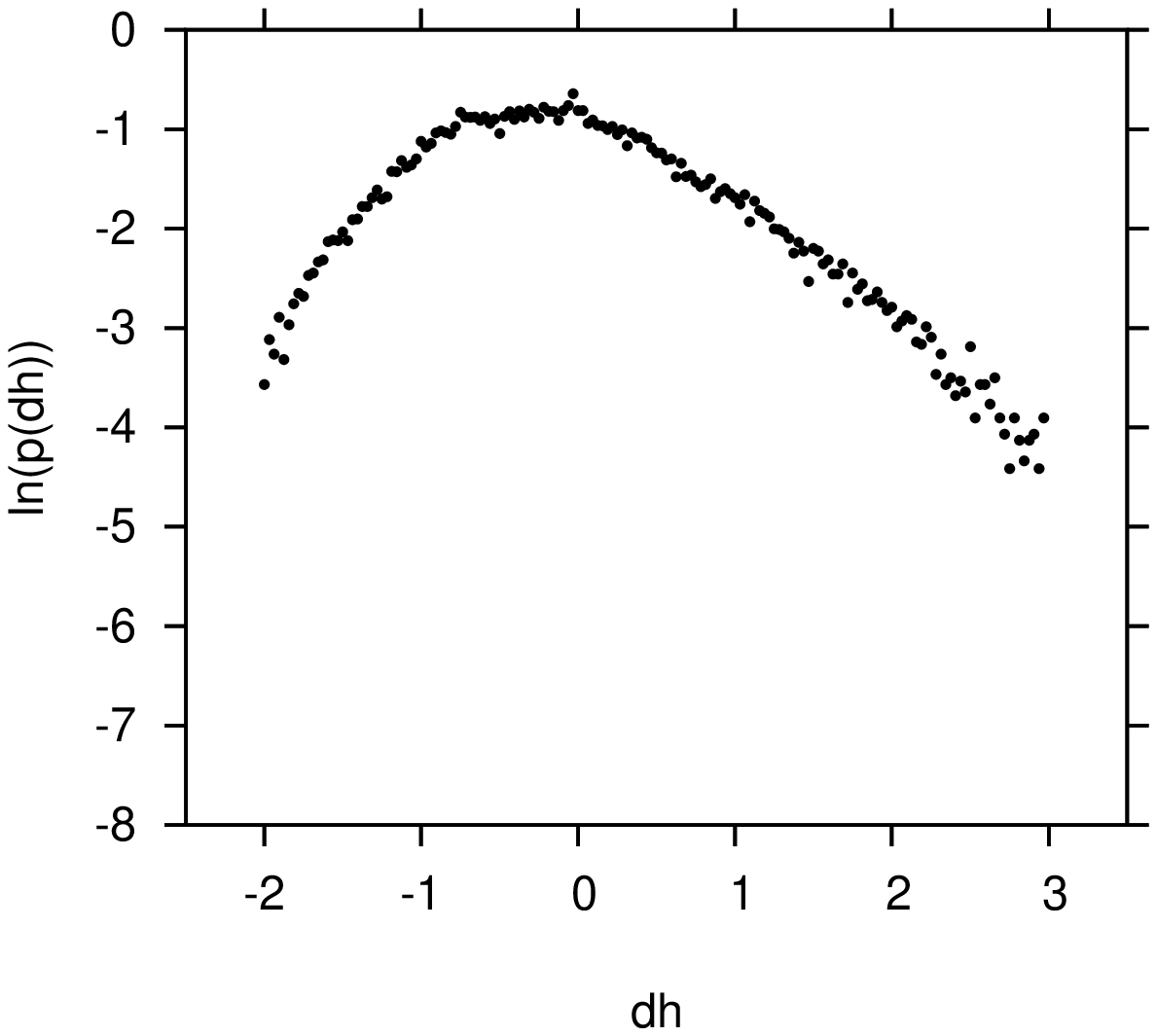}
\caption{\footnotesize{Log histogram of the seasonally adjusted daily water height Danube river data.}}
\label{fig:donaubinlogdata}
\end{minipage}
\end{figure}
\noindent
In figures \ref{fig:donaubin3},\ref{fig:donaubinlogdata} we show the Histogram and Log histogram of $\sigma_hP(h)$ against $\left(\frac{h-\overline{h}}{\sigma_h}\right)$. 

\begin{figure}[htb]
\begin{center}
\includegraphics*[width=8cm]{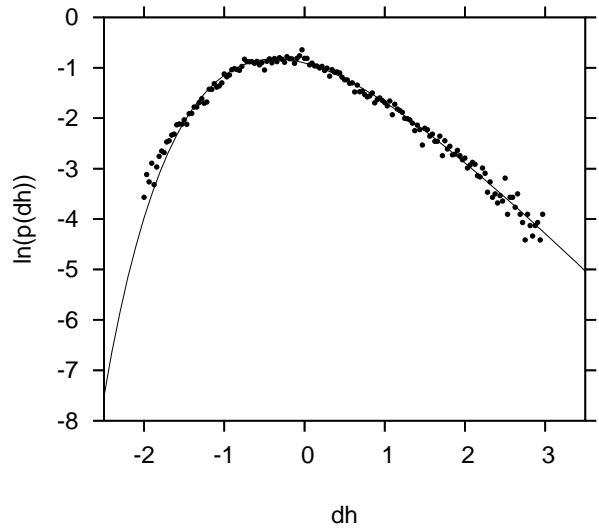}
\caption{\footnotesize{Logarithm of the reversed BHP PDF on top of the Log histogram of the Danube seasonally adjusted data. Near the origin there is less deviation from the BHP. The left tail of the Log histogram seems to follow a different asymptote than that of the BHP. The right tail the of the BHP asymptote seems to be a good fit and the higher dispersion around the BHP curve may be due to measurement error because the values used in that section are the height maximums.}} \label{fig:donaubinlog3}
\end{center}
\end{figure}
\noindent
As it may seen in figure \ref{fig:donaubinlog3} the reverse Log-BHP falls on top of the Log-histogram of the deseasonalised Danube data.\\
The analogue histograms for the Douro river show a higher concentration of values below the origin (mean for the original data). This is not surprising because for the Douro basin there is no melting ice to feed the river in the spring and in the long dry summer where the natural flow is much less than that of winter. The mean daily flow of Douro is close to the percentile 70. For the Danube data the average height is about the same as the median.
\begin{figure}[!htb]
\begin{minipage}[t]{0.45\linewidth}
\includegraphics[width=\linewidth]{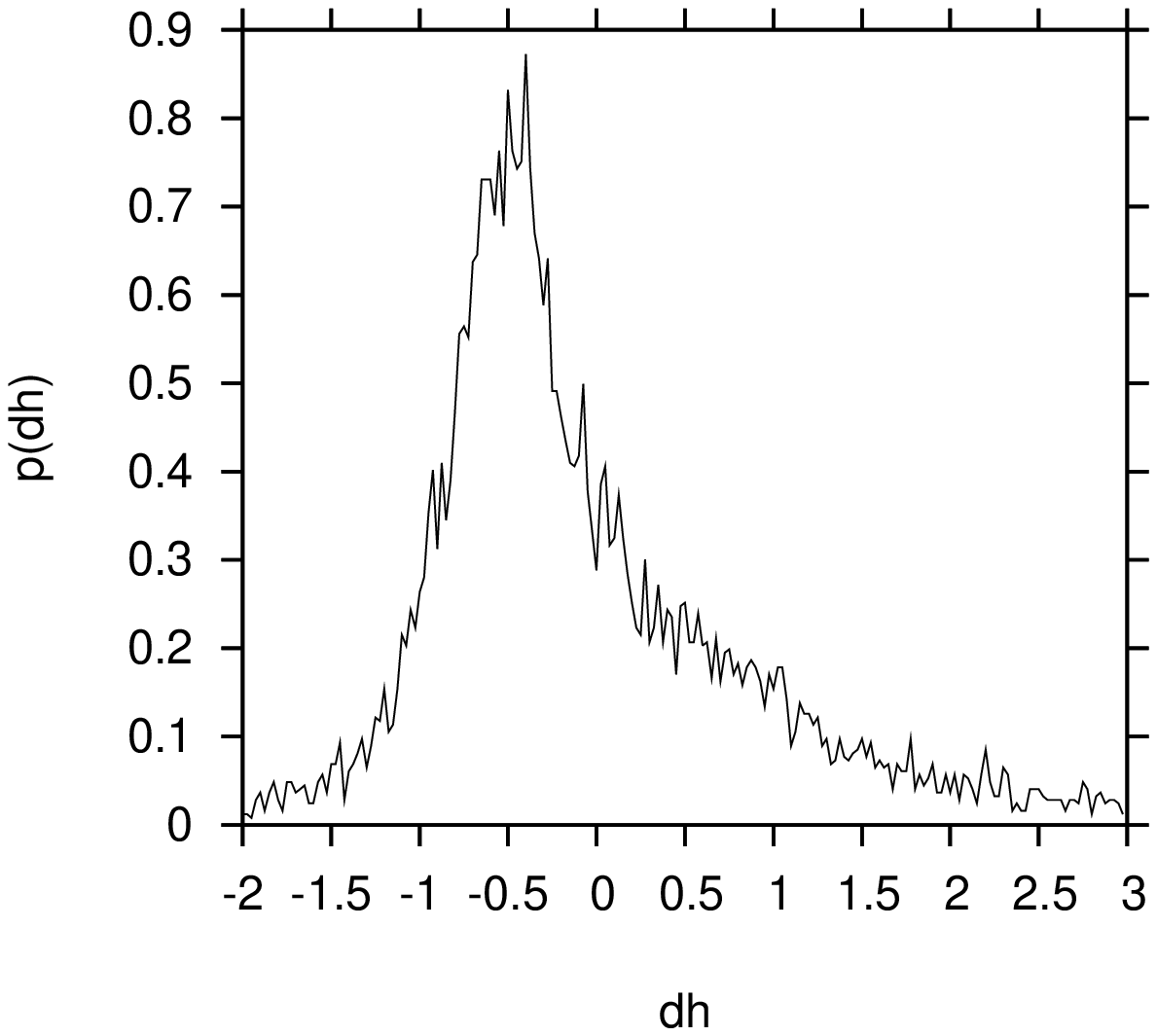}
\caption{\footnotesize{Histogram of the seasonally adjusted daily water height Douro river data.}}
\label{fig:dourobinnew}
\end{minipage} \hfill
\begin{minipage}[t]{0.45\linewidth}
\includegraphics[width=\linewidth]{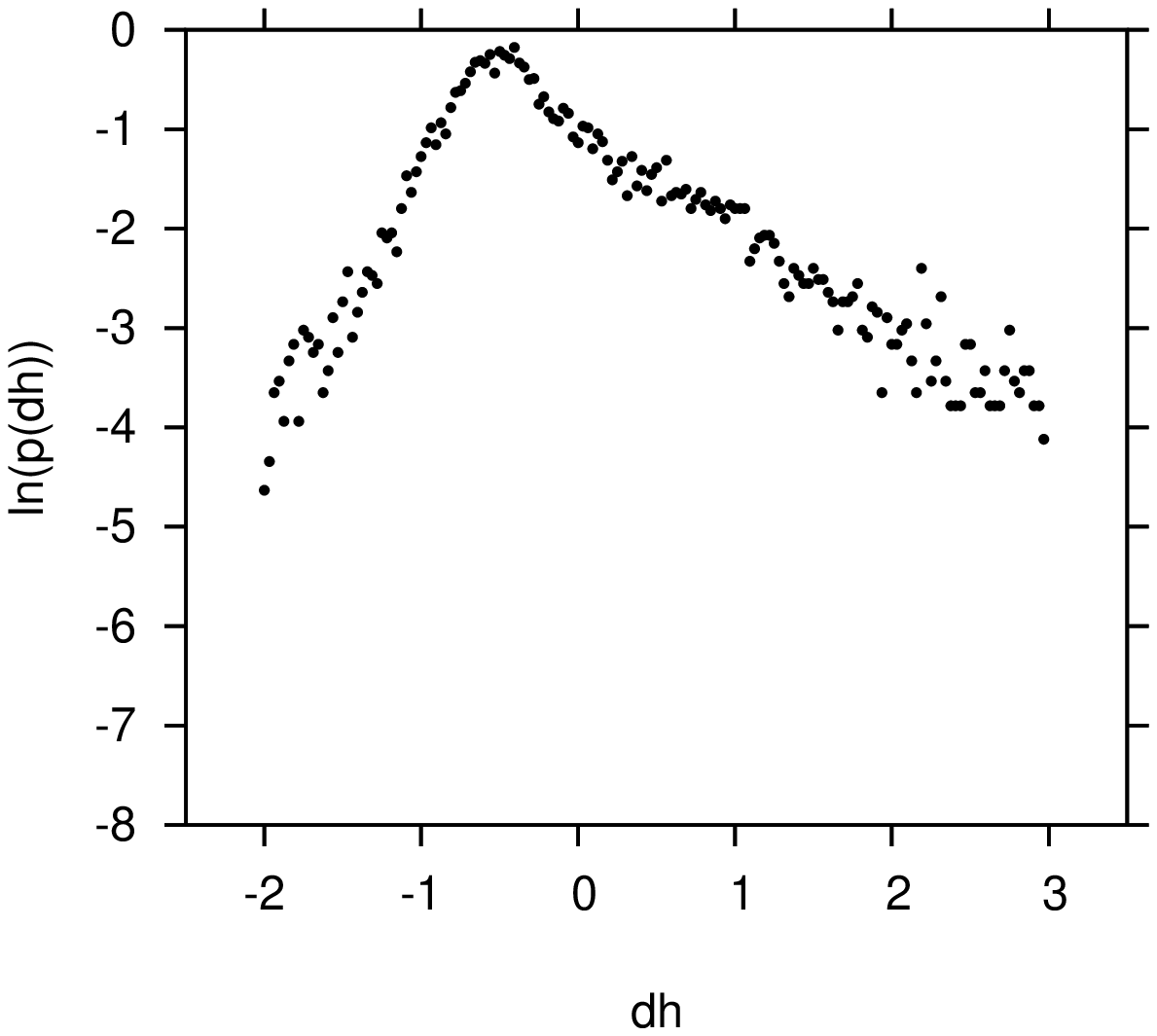}
\caption{\footnotesize{Log histogram of the seasonally adjusted daily water runoff of the Douro river data. The spike at the origin (mean of the original data) is explained by the high frequency of small runoffs.}}
\label{fig:dourobinlogdata}
\end{minipage}
\end{figure}
\noindent
The predominance of small runoffs for the Douro may be seen in the deseasonalised histogram and Log histogram, figures \ref{fig:dourobinnew},\ref{fig:dourobinlogdata}. In this last graphic the differences for the Danube are made very clear because of the spike at the origin.

\begin{figure}[!htb]
\begin{minipage}[t]{0.45\linewidth}
\includegraphics[width=\linewidth]{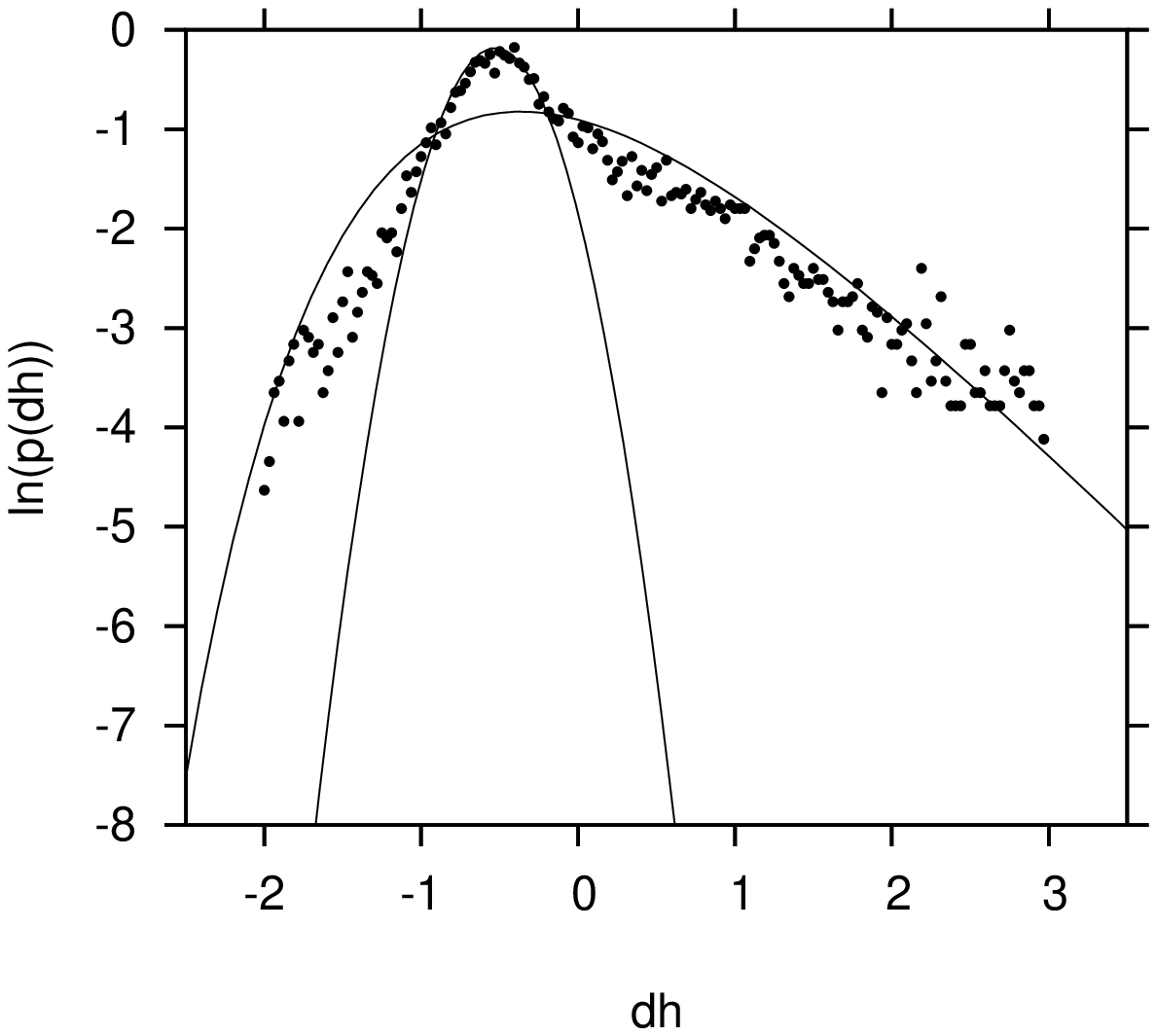}
\caption{\footnotesize{Log Histogram of the seasonally adjusted Douro river daily water level, for the full year with the BHP and the Gaussian fit on top. The deviations from the BHP at the origin are well fitted with the Gaussian PDF. This fact may be due to regulation measures at the dams. For smaller and specially for larger runoffs the seasonally adjusted flow tend to the BHP PDF. }}
\label{fig:dourobinloggauss}
\end{minipage} \hfill
\begin{minipage}[t]{0.45\linewidth}
\includegraphics[width=\linewidth]{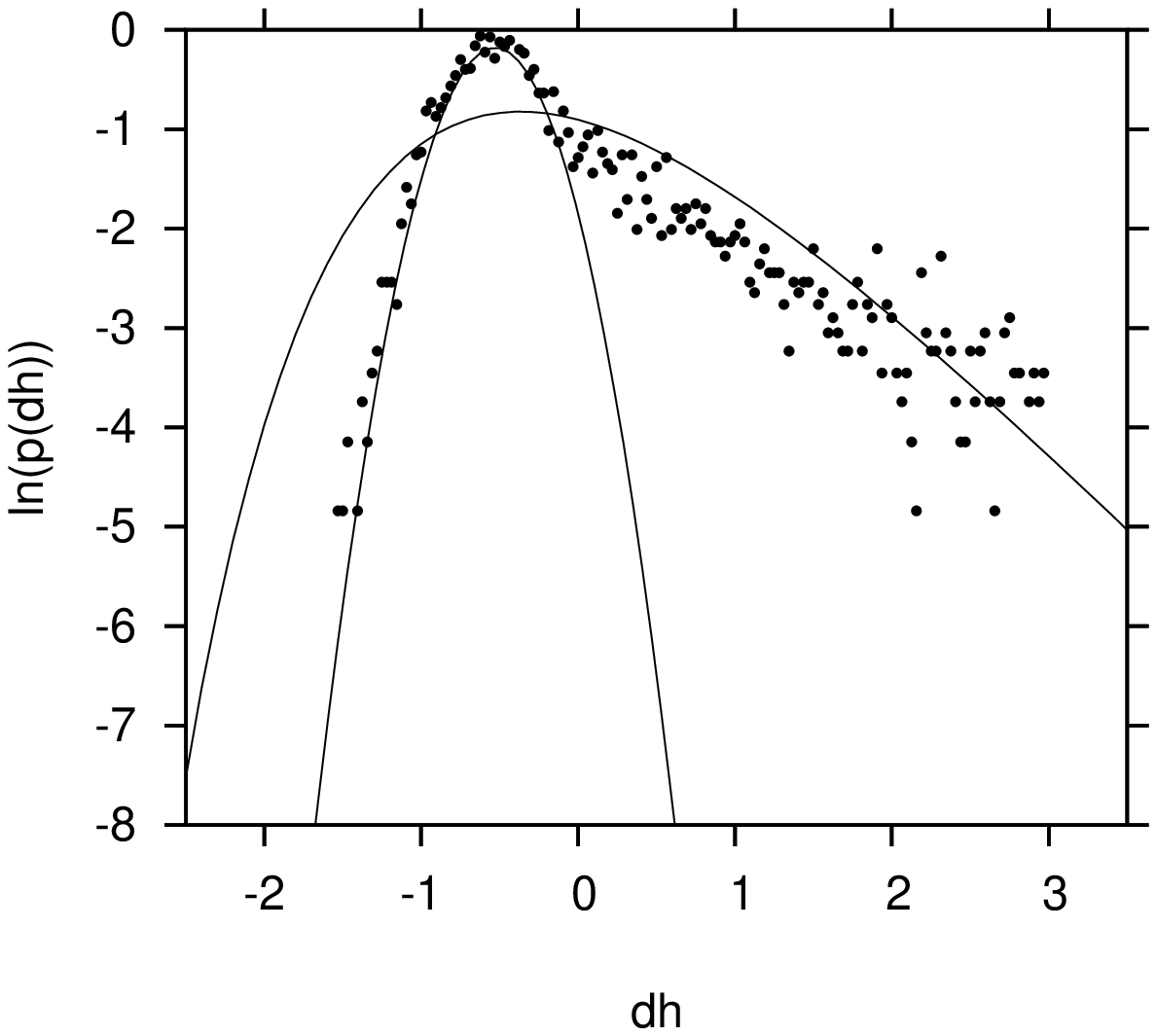}
\caption{\footnotesize{Log Histogram of the seasonally adjusted Douro river daily water level, autumn and winter with the BHP and the Gaussian fit on top. In these seasons the smaller runoffs can be better regulated towards the Gaussian PDF and larger runoffs tend to occur more frequently which explains the larger number of points around the BHP curve. The resulting PDF is a mixture of a gaussian up to a certain control threshold and the BHP for larger runoffs.}}
\label{fig:dourobinloggausswinter}
\end{minipage}
\end{figure}
\noindent
Finally, in figure \ref{fig:dourobinloggauss} is shown the Log of the BHP and the Gaussian fit\footnote{only a subset of data was used for the fit.} on top of the Log histogram of the deseasonalised daily runoff data for the full year and only for the winter regime. The BHP PDF comes from a system with no outside tuning so it should model data that is very close to the natural flow of the river. The river flow is more close to its natural form when regulation is not applicable. This is the case when the river carries so much water that the dams have to be set open, see figures \ref{fig:dourobinloggauss} and \ref{fig:dourobinloggausswinter}. The differences in the left branch of figures \ref{fig:dourobinloggauss} and \ref{fig:dourobinloggausswinter} appears to be caused by a regulation for small Winter runoffs which seems not to be possible or desirable for the summer resulting in a bad fit for the full year. The PDF of the deseasonalised daily Douro runoff data in Winter results is a mixture of a Gaussian PDF for small and medium runoffs and BHP for runoffs above a certain threshold. Future work passes by searching for stronger evidence in favour or against the allegations concerning the differences between the so called \textit{universal rivers} and those which seem not to be universal and also the differences between the winter regimes.\\
The BHP PDF has been found to be an explanatory model for fluctuations of several phenomenon such as, width power in steady state systems, variations in river heights and flow, numerous self-organised critical systems among others, (see references in \cite{BanksBramwell}). Still it is not known yet why BHP captures the essential behaviour of systems of different universality classes. Finally there could be a relation between the Fisher-Tippett-Gumbel distribution of extremal statistics and the BHP but recent works \cite{Bramwelletal2001,DahlstedtJensen2001}, showed asymptotic differences between the two so in this direction there is also work to be done.

\section*{Acknowledgements}

We thank Imre J\'anosi for providing the Danube data and him, Jason Gallas and Henrik Jensen for discussions and literature.




\biblio{

\bibitem{GonStoPin06}
      Gon\c calves, R., Stollenwerk, N. and Pinto, A.
      An\'alise de scaling em rios e a hip\'otese de universalidade da distribui\c c\~ao BHP,
      {\it Actas do XIII Congresso Anual da Sociedade Portuguesa de Estatistica.}, 389--400.

\bibitem{ImreJason99}
        J\'anosi, I.M., \& Gallas, J.A.C. (1999) 
        Growth of companies and water-level fluctuations of the
        River Danube,
        {\it Physica }{\bf A 271}, 448--457.


\bibitem{DahlstedtJensen2005}
        Dahlstedt, K., \& Jensen, H.J. (2005) 
        Fluctuation spectrum and size scaling of river flow and level,
        {\it Physica }{\bf A 348}, 596--610.


\bibitem{DahlstedtJensen2001}
        Dahlstedt, K., \& Jensen, H.J. (2001) 
        Universal fluctuations and extreme-value statistics,
        {\it J. Phys. A: Math. Gen. }{\bf 34}, 11193--11200.

\bibitem{Bramwelletal2001}
        Bramwell, S.T., Fortin, J.Y., Holdsworth, P.C.W., 
        Peysson, S., Pinton, J.F., Portelli, B. \& Sellitto,M. (2001) 
        Magnetic Fluctuations in the classical XY model:
        the origin of an exponential tail in a complex system,
        {\it Phys. Rev  }{\bf E 63}, 041106.
        Also as {\it cond-mat/0008093v2, 12. Dec. 2000}.


\bibitem{bramwellfennelleurophys2002}
        Bramwell, S.T., Fennell, T., Holdsworth, P.C.W.,
        \& Portelli, B. (2002) Universal Fluctuations of the Danube Water
        Level: a Link with Turbulence, Criticality and Company Growth,
        {\it Europhysics Letters  }{\bf 57}, 310.
        Also as {\it cond-mat/0109117v2, 8. Sep. 2001}.


\bibitem{bramwell2000}
        Bramwell, S.T., Christensen, K., Fortin, J.Y., Holdsworth, P.C.W.,
        Jensen, H.J., Lise, S., L\'opez, J.M., Nicodemi, M.
        \& Sellitto,M. (2000) 
        Universal Fluctuations in Correlated Systems,
        {\it Phys. Rev. Lett. }{\bf 84}, 3744--3747.


\bibitem{BHP1998}
        Bramwell, S.T., Holdsworth, P.C.W., \&  Pinton, J.F. (1998)
        Universality of rare fluctuations in turbulence and critical
        phenomena
        {\it Nature }{\bf 396}, 552--554.


\bibitem{Archambault1997}
        Archambault, P., Bramwell, S.T., \& Holdsworth, P.C.W. (1998) 
        Magnetic fluctuations in a finite two-dimensional XY model,
        {\it J. Phys. A: Math. Gen. } {\bf 30}, 8363--8378.


\bibitem{BTW88}
        Bak, P., Tang, C., \& Wiesenfeld, K. (1988) Self-organized criticality.
        {\it Phys. Rev. }{\bf A 38}, 364--374.


\bibitem{BanksBramwell}
        Banks, S. T., \& Bramwell, S. T.  (2005) 
        Temperature dependent fluctuations in the two-dimensional \textit{XY} model ,
        {\it cond-mat/0507424v1,18 July 2005}.

\bibitem{RayAguaJensen}
 Sinha-Ray, P., Borda \'Agua, L., Jensen, H. J. (2001) 
 Threshold dynamics, multifractality and universal fluctuations in the SOC forest-fire: Facets of an auto-ignition model,
  {\it Physica D} {\bf 157 3}, 186 - 196.

\bibitem{Stanleyetal}
        Stanley, M.H.R., Amaral, L. A. N., Buildrev, S. V., Havlin, S., Leschhorn, H.
        Maass, P., Salinger, M.A. \& Stanley, E. (1996)
        Scaling behaviour in the growth of companies.
        {\it Letters to Nature} {\bf 379}, 29, 804-806.
        
        
\bibitem{MackPalmaVergara2005}
        Mack, G., Palma, G., \& Vergara, L. (2005) 
        Corrections to Universal Fluctuations in Correlated Systems:
        the 2D XY-model,
        {\it cond-mat/0506713v1,28 June 2005}.
        
\bibitem{Lenz}
        Lenz, W. (1920)
        Beitrag zum Verständnis der magnetischen Eigenschaften in festen Körpern.        
        {\it Phys. Zeitschr.} {\bf 21}, 613-615.
        
\bibitem{Ising}
        Ising, E. (1925)
        Beitrag zur Theorie des Ferromagnetismus.
        {\it Zeitschrift f\"ur Physik }{\bf 31}, 253-258.
        
\bibitem{Onsager}
          Onsager, L. (1944)
          Crystal Statistics. I. A Two-Dimensional Model with a Order-Disorder Transition.
           {\it Phys. Rev.} {\bf 65}, 117-149.
        
\bibitem{KosterlitzThouless} 
        Kosterlitz, J M. \& Thouless, J. (1973)
        Ordering, metastability and phase transitions in two-dimensional systems.
        {\it J. Phys. C.}{\bf 6}, 1181-1203.                 

\bibitem{Binneyetal} 
        Binney, J.J., Dowrick, N.J., Fisher, A.J., \& Newman, M.E.J. (1992). 
        {\it The Theory of Critical Phenomena, An Introduction to the
        Renormalization Group}
        (Oxford University Press, Oxford).

\bibitem{Binder} 
        Binder, K. (1992). 
        {\it Computational Methods in Field Theory}
        (Springer, Berlin).
        
\bibitem{Yeomans} 
        Yeomans, J.M. (1992). {\it Statistical Mechanics of Phase Transitions}
        (Oxford University Press, Oxford).

\bibitem{ZinnJustin} 
        Zinn-Justin, J. (1989). {\it Quantum Field Theory and 
        critical phenomena}
        (Oxford University Press, Oxford).

}
\end{document}